\documentclass[aps,pre,
onecolumn,
groupedaddress,showpacs]{revtex4-2}

\usepackage{epsfig}
\usepackage{epstopdf}

\usepackage{graphicx}
\usepackage{xcolor}
\usepackage{amsmath}
\usepackage[T1]{fontenc}
\usepackage{caption}
\usepackage{float} 
\usepackage{dcolumn}
\usepackage{hyperref}
\usepackage{amsthm}
\usepackage{amssymb}
\usepackage[hyphenbreaks]{breakurl}
\def\d{\mbox{d}}

\begin{document}
\title{paper}
\author{ Petar Mali$^{1}$}  \author{Slobodan Rado\v sevi\' c $^{1}$} \author{Sonja Gombar$^{1}$} \author{Jasmina Teki\' c $^{2}$} \author{Milan Panti\' c $^{1}$} \author{Milica Pavkov-Hrvojevi\' c $^{1}$}
\affiliation{$^1$ Department of Physics, Faculty of Sciences, University of Novi Sad,
Trg Dositeja Obradovi\' ca 4, 21000 Novi Sad, Serbia}
\affiliation{$^2$ "Vin\v ca" Institute  of Nuclear Sciences, Laboratory for Theoretical and Condensed Matter Physics - 020, University of Belgrade, PO Box 522, 11001 Belgrade, Serbia}


\title{The largest Lyapunov exponent as a tool for detecting relative changes in the particle positions}

\date{\today}


\begin{abstract}

Dynamics of the driven Frenkel-Kontorova model with asymmetric deformable substrate potential is examined by analyzing response function, the largest Lyapunov exponent and Poincar\'{e} sections for two neighboring particles. The obtained results show that the largest Lyapunov exponent, besides being used for investigating integral quantities, can be used for detecting microchanges in chain configuration of both damped Frenkel-Kontorova model with inertial term and its strictly overdamped limit. Slight changes in relative positions of the particles are registered through jumps of the largest Lyapunov exponent in the pinning regime. The occurrence of such jumps is highly dependent on type of commensurate structure and deformation of substrate potential. The obtained results also show that the minimal force required to initiate collective motion of the chain is not dependent on the number of Lyapunov exponent jumps in the pinning regime. These jumps are also registered in the sliding regime, where they are a consequence of a more complex structure of largest Lyapunov exponent on the step.

\end{abstract}
 \maketitle

\section{Introduction}
\label{intro}

Synchronization effects have been a subject of intensive theoretical and experimental studies in charge density wave transport \cite{Zettl,Gru,Grun,Gruner,Thorne,Thorne2,Thorne3}, vortex matter \cite{Vinokur,Reichhardt,Kol}, irradiated Josephson junctions \cite{Benz,Sell,Free,Shukrinov,Shukrinov2}, superconducting nanowires \cite{Dins,Bae} and driven colloidal systems \cite{Nature}. One of the models used for investigation of synchronization phenomena  (mode-locking, i.e. Shapiro steps), starting from microscopic dynamics, is the dissipative Frenkel-Kontorova (FK) model under external periodic forces \cite{monografija,Flor,Falo, FlorAP}. Recently, it has been shown that dc-driven FK model with a lateral periodic excitation to the substrate potential can be used to generate Shapiro steps as well \cite{weilei}. The one-dimensional FK model represents a chain of coupled particles which are subjected to the substrate potential. For the standard FK model, the particles are harmonically coupled to their nearest neighbors and the substrate potential is sinusoidal. To capture certain phenomena in Josephson junction arrays \cite{Shukrinov3,unharmonic,unharmonic2}, charge density wave systems \cite{Thorne,Thorne2,Thorne3} and tribology \cite{tribology1,tribology2,tribology3} different generalizations of the FK model are used.

Due to competition between the length scales of interparticle and substrate potential, many nontrivial ground states of the model are possible. They can be classified into two categories - commensurate (for which the interparticle average distance, i.e. winding number, is rational) or incommensurate (for which the winding number is irrational) \cite{monografija,FlorAP,OBBook}. If an external dc driving force is
applied, there exists a critical threshold value, i.e. critical depinning force $F_{\rm{c}}$,  which separates two dynamical regimes - pinning and
sliding regime. The latter regime is defined by collective motion of the particles along with non-zero average velocity, whereas in the pinning regime the particles are pinned to the static configurations with zero average velocity. When both external dc and ac forces are applied to the FK model, mode-locking appears due to locking between the frequency of particle motion and frequency of external ac force, which results in the staircase-like response function \cite{Falo,hu1,hu2,Tekic1,Tekic2,Tekic3,Mali}. A new class of nonlinear
periodic deformable potentials, which could be specified by suitable choice of parameters, was introduced in \cite{miseli,miseli2}. By fixing parameters, different periodic potentials, such as assymetric deformable potential, which are relevant for studying Josephson junctions, charge density waves, heat conduction in low dimensional lattices, and crystals with dislocations, can be obtained. Unlike the case of the standard FK model, when the subharmonic steps either do not exist or they are too small, for the FK model with asymmetric deformable substrate potential large subharmonic steps can be observed in the response function plot \cite{hu1,hu2,Tekic2}.

Aside from the response function, the largest Lyapunov exponent (LE) also provides certain insight into the dynamics of dissipatively driven FK model. It has been shown that the largest LE can be viewed as a tool to characterize chaotic, periodic, and quasiperiodic motion (see \cite{politi,randommat} and references therein). Furthermore, the LEs are also used for observation of dynamical phase transitions (pinning to sliding regime and unlocking transition) \cite{Falo,application,Sokolovic,Shukrinov,Shukrinov2} and thermally induced phase transitions \cite{pt1,pt2,pt3,pt4}. Specifically, the largest LE calculated for the standard overdamped FK model in the sliding regime takes on negative values on the steps, determining the trajectories periodic in time, while it reaches zero outside the steps, characterizing the unlocking transition to the quasiperiodic regime \cite{Falo,application}. On the other hand, in the underdamped regime chaos sometimes appears at the edges of the Shapiro steps and system exhibits structural chaotic behavior \cite{Shukrinov2,weilei}. Moreover, according to \cite{farey} calculating the largest LE presents the most sensitive way for detection of the Shapiro steps, especially the subharmonic ones. Also, it has recently been shown that the LE spacings for certain parameters can change from Poissonian statistics to Wigner’s surmise \cite{randommat}. In the case of Wigner's surmise the largest LE fluctuation statistics shows behavior similar to the Tracy-Widom  distribution \cite{randommat}. 

In the present paper, we will show that the  largest LE can be applied to investigate the changes in relative positions of the particles in generalized dissipatively driven FK model with asymmetric deformable substrate potential. It will be shown that calculation of the largest LE can be useful for analysis of dynamics of ac+dc driven FK model even in the pinning regime. In order to confirm the nature of the structural changes in chain configuration, we will compare the results obtained by the largest LE analysis with Poincar\'{e} sections for two neighboring particles. The paper is organized as follows: The model is introduced in Sec. II and simulation results are presented in Sec. III. Finally, Sec. IV concludes the paper.

\section{Model}
\label{model}

In this paper, we investigate both strictly overdamped and damped generalized Frenkel-Kontorova model with asymmetric deformable potential (ASDP):
\begin{equation}
V(u)=\frac{K}{(2\pi)^{2}}\,\,\frac{\big(1-r^{2}\big)^{2}\big(1-\cos(2\pi u)\big)}{\big[1+r^{2}+2r\cos(\pi u)\big]^2},
\end{equation}
where $r$ is the shape parameter and $K$ pinning strength, driven by periodic force:
\begin{equation}
F(t)=F_{\mathrm{dc}}+F_{\mathrm{ac}}\cos(2\pi\nu_{0}t),
\end{equation}
where $F_{\mathrm{dc}}$ is the dc force and $F_{\mathrm{ac}}$ and $\nu_{0}$ are the amplitude and frequency of the ac force, respectively. Notice that the choice of $r=0$ corresponds to the simple sinusoidal substrate potential of the standard FK model, whereas it is deformable for $0<|r|<1$. The system of the equations of motion for strictly overdamped FK model is given by $N$ first order nonlinear differential-difference equations:
\begin{equation}
\dot{u}_{l}=u_{l+1}+u_{l-1}-2u_{l}-V'(u_{l})+F(t), \hspace{4mm} l=1,2,...,N \label{eqmot}
\end{equation}
with $N$ being the number of particles and $u_{l}$ the position of the $l$-th particle. 

\begin{figure}[h!]
\centering
\includegraphics[scale=0.63]{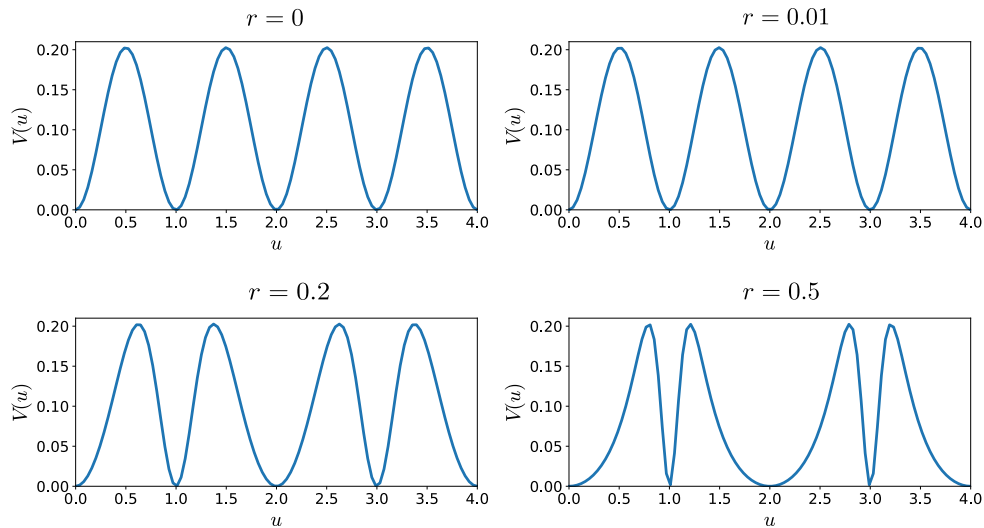}
\caption{Asymmetric deformable potential for $K=4$ and four different
values of the shape parameter $r$.}
\label{fig:potential}
\end{figure}

The competition between two frequency scales (one of the external periodic force and the other of the motion of particles over the sinusoidal potential driven by $F_{\rm{dc}}$) results in the appearance of Shapiro steps. These steps correspond to the resonant solutions of \eqref{eqmot}. If $\lbrace u_{l}(t)\rbrace$ is the solution of \eqref{eqmot} with initial condition $\lbrace u_{l}(t_{0})\rbrace$, then:
\begin{equation}
\sigma_{i,j,m}\lbrace u_{l}(t)\rbrace=\lbrace u_{l+i}(t-\frac{m}{\nu_{0}})+ja\rbrace
\end{equation}
is also a solution of the same equations corresponding to the initial condition $\sigma_{i,j,m}\lbrace u_{l}(t_{0})\rbrace$. Set of indices $i$, $j$ and $m$ are arbitrary integers with $i$ defining relabeling of the particles and $j$ and $m$ corresponding to space and time translations, respectively, whereas $a$ is the periodicity of the potential $V(u)$ ($a=1$ for undeformed case corresponding to $r=0$ and $a=2$ when deformation is included). The average velocity of the resonant solution satisfies the equation:
\begin{equation}
\bar{v}=\frac{i\omega+ja}{m}\nu_{0},
\end{equation} 
where $\omega$ is the winding number, i.e. interparticle average distance \cite{farey}. In the case of commensurate structures we consider in the present paper, $\omega=\frac{p}{q}$, where $p$ and $q$ are coprime integers, meaning that $q$ particles are distributed over $p$ substrate potential wells in average. For $m=1$, the steps are called harmonic, whereas for $m>1$ they are subharmonic. 

\begin{figure}[h!]
\centering
\includegraphics[scale=0.63]{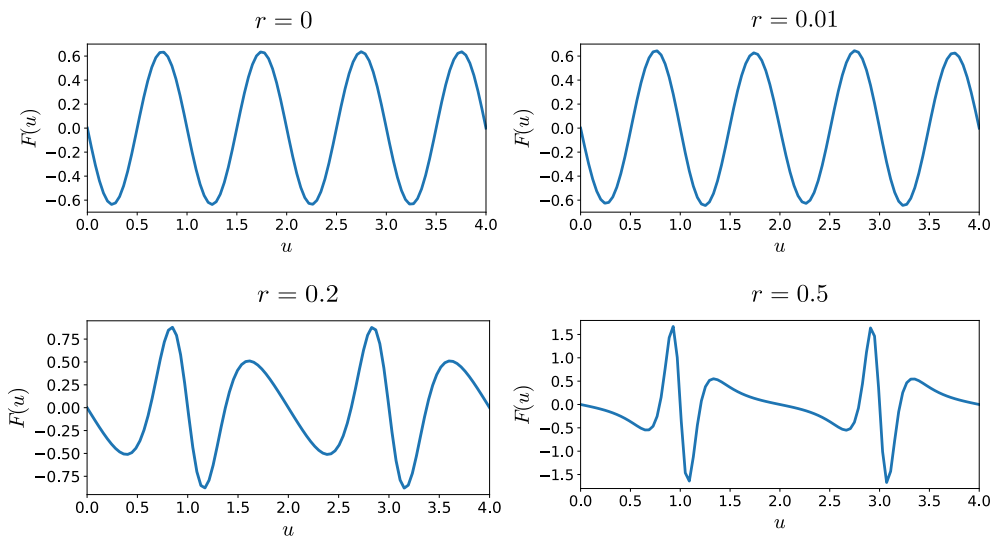}
\caption{Pinning force $F(u)=-V'(u)$ for $K=4$ and four different values of the shape parameter $r$.}
\label{fig:force}
\end{figure}

ASDP and the corresponding pinning force for $K=4$ and four values of shape parameter $r$ are presented at Figs. \ref{fig:potential} and \ref{fig:force}, respectively. By increasing the value of the shape parameter $r$, two nonequivalent types of potential wells appear (one with flat and the other with sharp bottom), which changes the periodicity of the substrate potential from $a=1$ (for $r=0$) to $a=2$ (for $0<|r|<1$) and effectively increases number of degrees of freedom \cite{Tekic1,saturation,farey,asdp}. In the case of $\omega=\frac{1}{q}$, where $q$ represents the average number of particles per a potential well, harmonic and subharmonic steps are numbered according to $\bar{v}=\frac{i\omega}{m}\nu_{0}$ \cite{farey,Sonja}.

To obtain response functions, equations of motion \eqref{eqmot} were integrated according to the fourth order Runge-Kutta method with periodic boundary conditions. Detailed explanation of the approach used to obtain the largest LE is given in \cite{sprot}. In order to calculate the largest LE a perturbed point $\tilde{u}_{l}$ is chosen as:
\begin{equation}
\tilde{u}_{l}(t_{\mathrm{ss}})=u_{l}(t_{\mathrm{ss}})\pm\sqrt{\frac{d_{0}^{2}}{N}},
\end{equation}
where $t_{\mathrm{ss}}$ is the necessary time for system to reach steady-state and $d_{0}=10^{-7}$ is the small parameter used to define perturbation of the initial configuration. The plus and minus sign in the equation appear with equal probability in order to reduce the possibility of projecting onto the subspace not dominated by the largest LE \cite{farey}. After steady states are reached for different values of dc force, the finite-time LEs are calculated, which for large values of $t$ converge to their asymptotic limit \cite{randommat}. 

\section{Results}
\label{results}

In this section, a comparative study of the generalized dissipatively driven FK model dynamics will be presented using the response function and the largest LE as a function of average driving force as well as the Poincar\'{e} sections for two neighboring particles.

\begin{figure}[h!]
\centering
\includegraphics[scale=0.78]{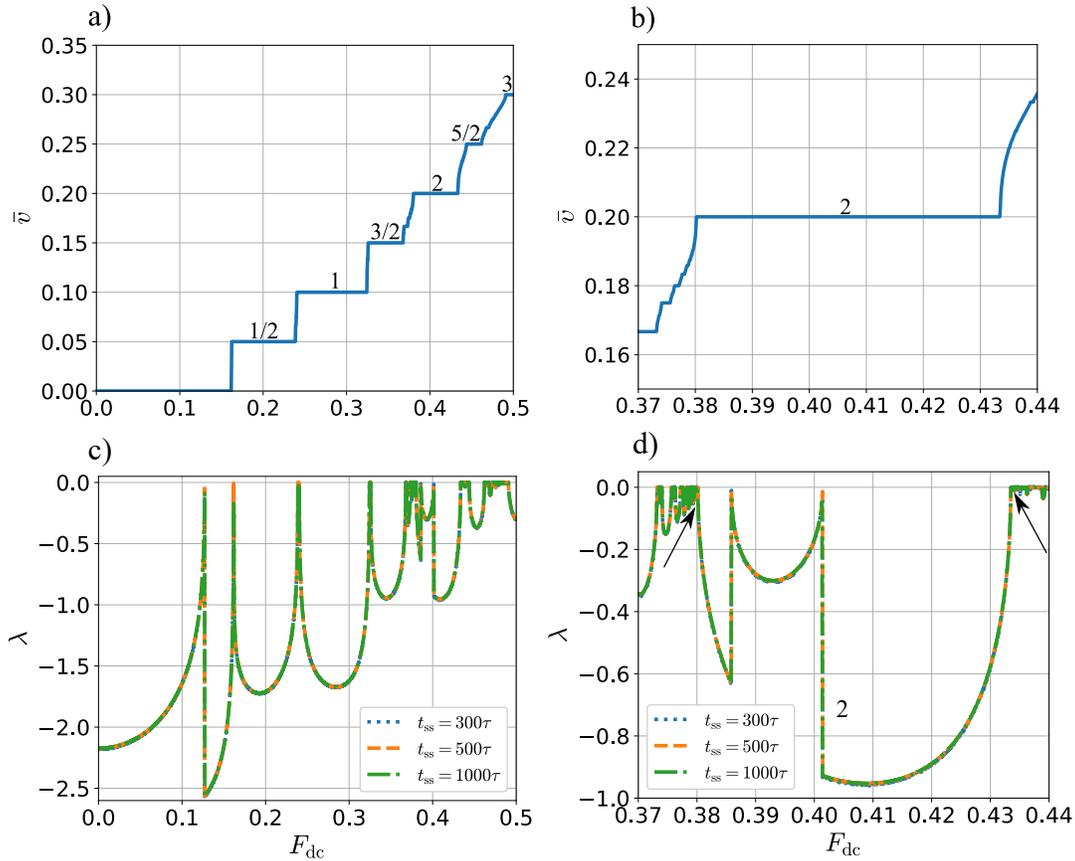}     
\caption{Average velocity a) and the largest LE c) as functions of dc force for $\omega=\frac{1}{2}$, $F_{\rm{ac}}=0.2$, $K=4.0$, $\nu_{0}=0.2$, $r=0.2$ and three different transient times $t_{\textrm{ss}}$. Zoomed parts of a) and c) are shown in b) and d), respectively. Large harmonic and subharmonic steps are marked as well. Arrows in d) mark the beginning and the end of the second harmonic step.} \label{fig:respfunler02}
\end{figure}

Average velocity \begin{equation}
\bar{v}=\big\langle\big\langle\dot{u}_{l}(t)\big\rangle\big\rangle_{T,N}=\lim_{T\to\infty}\frac{1}{TN}\sum_{l=1}^{N}\int_{t_{\mathrm{ss}}}^{t_{\mathrm{ss}}+T}\dot{u}_{l}(t)\d t,
\end{equation} and the largest LE are shown in Fig. \ref{fig:respfunler02} for wide range of dc force, which includes both pinning and sliding regime, in the case of $\omega=\frac{1}{2}$ (two particles per a substrate potential well in average) and $r=0.2$.  As mentioned before, for the generalized FK model with ASDP increasing the value of shape parameter $r$ effectively increases the number of degrees of freedom and hence large subharmonic steps are observed in Fig. \ref{fig:respfunler02} for $\omega=\frac{1}{2}$ and can also appear in any commensurate structure, even with integer values of winding number. Note that these steps are quite small in the case of the standard FK model with rational values of winding number, whereas for integer values of $\omega$ they do not exist since the overdamped FK model with integer value of $\omega$ reduces to single particle model \cite{farey,Sonja,hu1,rene, waldram}. Similarly to the standard FK model \cite{application}, the largest LE can be used for detection of both harmonic and subharmonic Shapiro steps and also dynamical phase transition from pinning to sliding regime. However, for the generalized FK model used in this paper the largest LE on the step shows a more complex structure as seen in Fig. \ref{fig:respfunler02} d). Moreover, an additional feature is observed in this case. The jump of the largest LE in pinning regime, in interval $F_{\textrm{dc}}\in(0.12159,0.1618)$ with minimum at $F_{\textrm{dc}}\approx 0.12701$, is seen in Fig. \ref{fig:respfunler02} c). The largest LE is calculated for three different transient times in order to verify that this jump is not an artifact of numerical calculations. Since the shape and the position of the jump remain the same as transient time is increased, we conclude that its appearance is a consequence of the system's dynamics. Furthermore, in the Appendix we show that the jump is not a result of the strictly overdamped limit as it is detected even when Wolf's algorithm for the damped FK model with inertial term is employed. It is thus necessary to further investigate the origin of this feature.   

\begin{figure}[h!]
\includegraphics[scale=0.9]{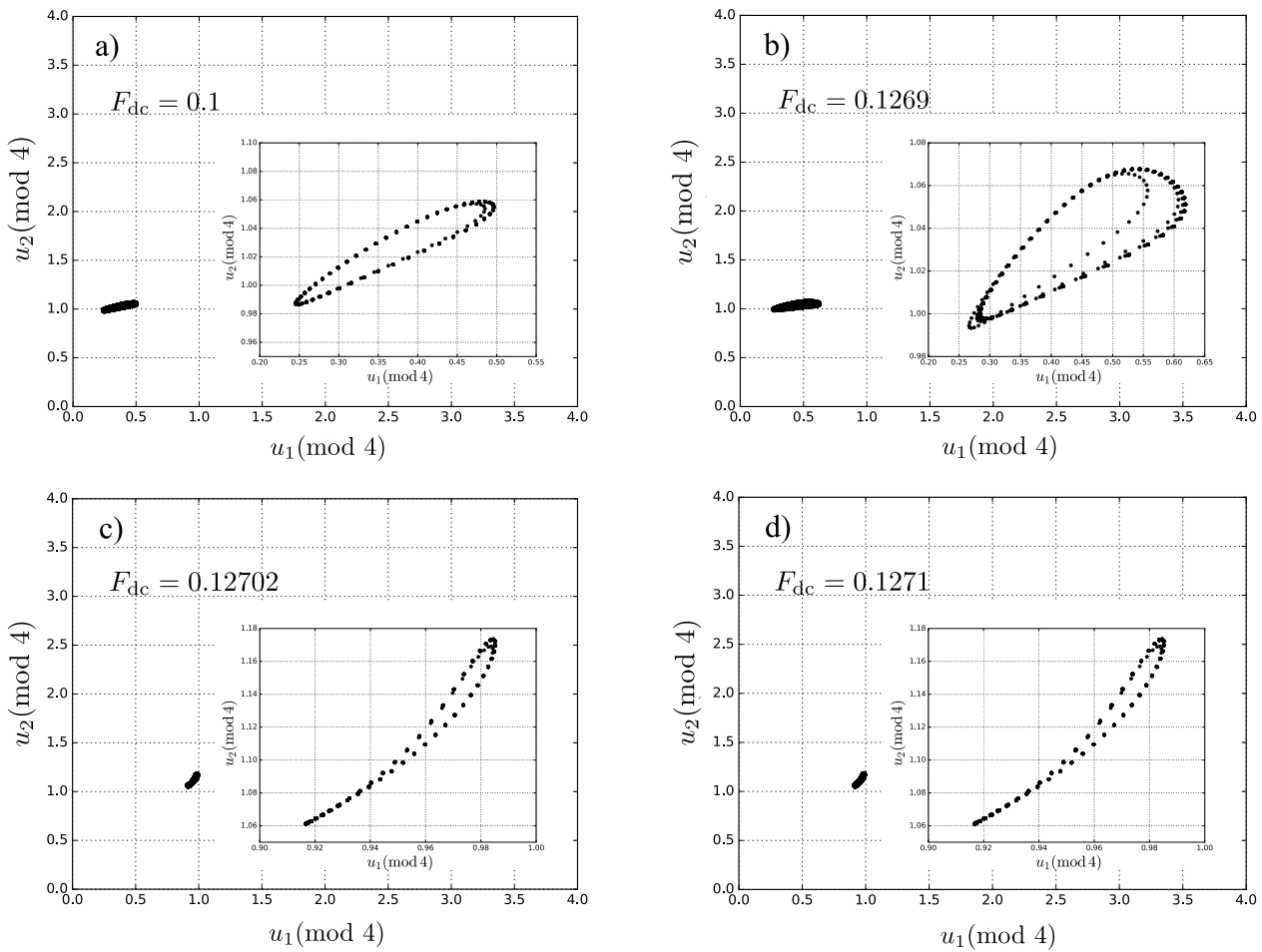} 
\centering
\caption{Poincar\'{e} sections for two neighboring particles with coordinates $u_{1}$ and $u_{2}$ for the given set of parameters: $\omega=\frac{1}{2}$, $F_{\rm{ac}}=0.2$, $K=4.0$, $\nu_{0}=0.2$ and $r=0.2$. The insets provide the enlarged view of Poincar\'{e} sections}. \label{fig:poincsec1}
\end{figure}

Due to the fact that the largest LE jump was observed in the pinning regime, where the average velocity is zero, logical framework of the investigation is to examine relative positions of the neighboring particles. In order to provide better visualization of the relative particle positions for the initial conditions $u_{l}=l\omega$, $l=1,...,N$, we used modulo operation. The relative motion of two neighboring particles is presented in Fig. \ref{fig:poincsec1} for values of dc force before the jump (a) and b)) and after the jump (c) and d)). The black regions correspond to the particles' jiggling (forward and backward motion) induced by the ac force (the insets show these regions in greater detail). By comparing Figs. \ref{fig:poincsec1} a) and b) with c) and d), one can notice the significant shift of region of jiggling after the jump of the largest LE. Notice that this shift is a consequence of the additional force (besides the driving force) originating from the ASDP that acts on the first particle of the initial configuration (see Fig. \ref{fig:force}). It allows the said particle to fall into the first narrow potential well seen in Fig. \ref{fig:potential}. Consequently, the system goes through structural change in chain configuration, passing from one pinned configuration to another. This structural change is directly witnessed through the jump of the largest LE in the pinning regime. \textit{Therefore, we corroborate that the largest LE, besides for detection of chaotic, periodic and quasiperiodic motion, can also be an important tool for spotting slight changes of the relative positions between particles in the system.} 

\begin{figure}[h!]
\includegraphics[scale=0.9]{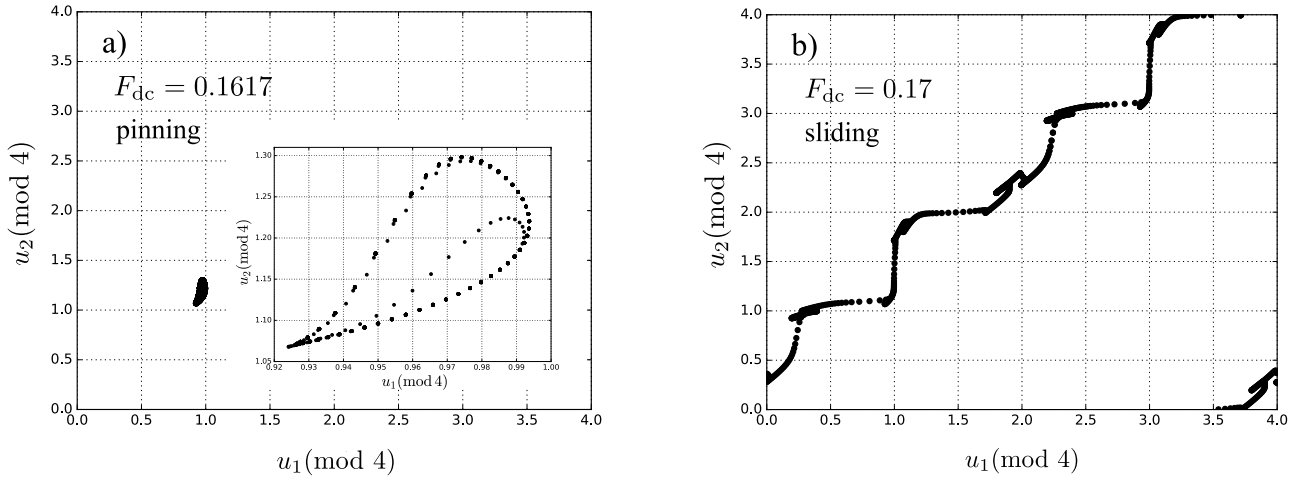} 
\centering
\caption{Poincar\'{e} sections for two neighboring particles with coordinates $u_{1}$ and $u_{2}$ for the given set of parameters: $\omega=\frac{1}{2}$, $F_{\rm{ac}}=0.2$, $K=4.0$, $\nu_{0}=0.2$ and $r=0.2$. The inset shows the zoomed part of the Poincar\'{e} section in the pinning regime.}  \label{fig:poincsec2}
\end{figure}

However, additional increase of dc force is required for the system to go through transition to sliding regime. In this concrete situation, the critical depinning force $F_{\rm{c}}$ is around $F_{\rm{dc}}\approx0.1618$. In Fig. \ref{fig:poincsec2} one can see the difference between the pinning a) and sliding b) regime. The latter case corresponds to the situation where the collective motion appears, which yields $\bar{v}\neq 0.$

\begin{figure}[h!]
\includegraphics[scale=0.65]{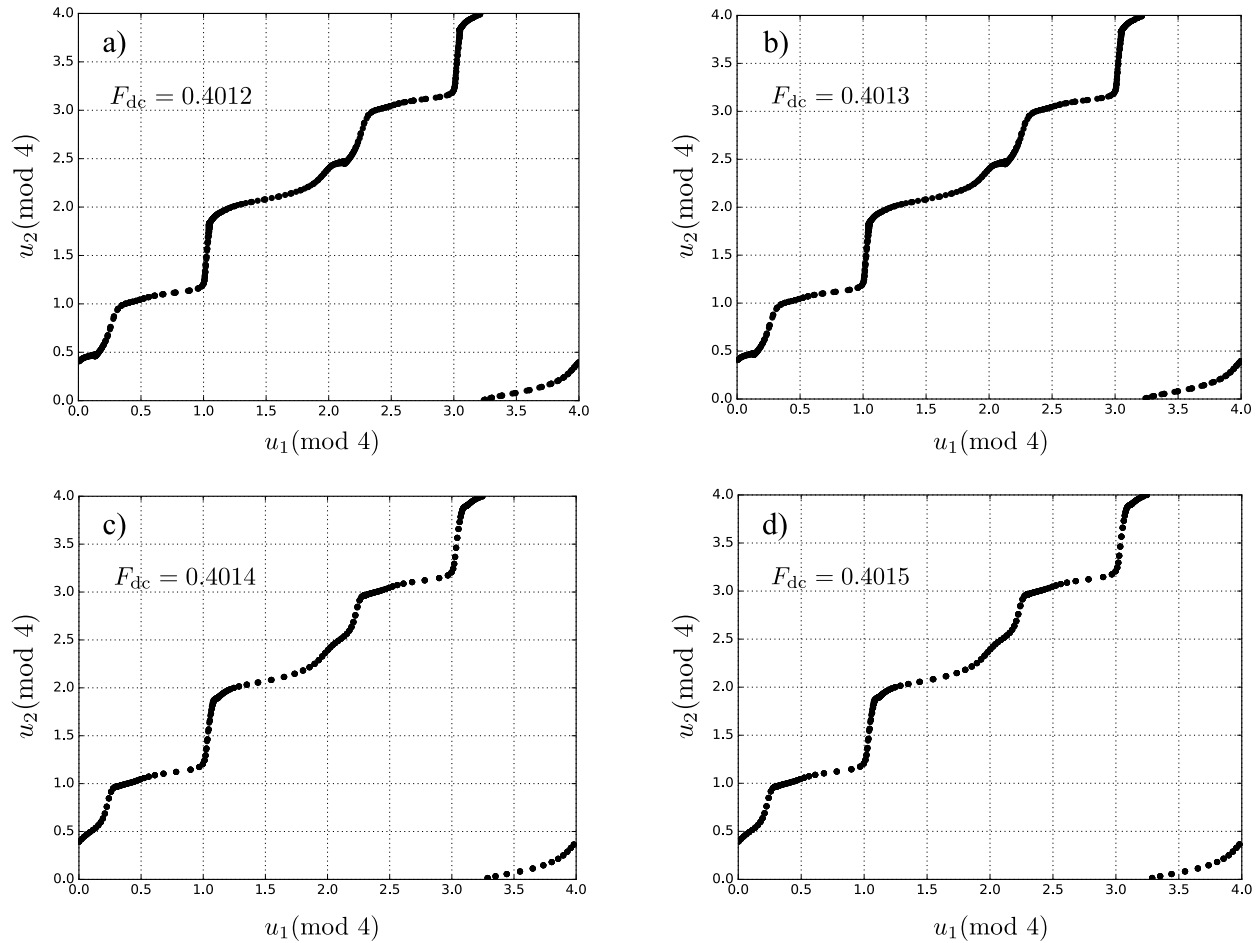} 
\centering
\caption{Poincar\'{e} sections for two neighboring particles with coordinates $u_{1}$ and $u_{2}$ for the given set of parameters: $\omega=\frac{1}{2}$, $F_{\rm{ac}}=0.2$, $K=4.0$, $\nu_{0}=0.2$ and $r=0.2$.} \label{fig:poincsecad}
\end{figure}

A similar shift to the one detected in the pinning regime by the largest LE jump also occurs in the sliding regime. This is the cause of the more complex structure on the step observed in Fig. \ref{fig:respfunler02} d). To prove that claim, we investigated the largest LE jump on the second harmonic step which occurs for $F_{\mathrm{dc}}=0.4014$. By examination of this using Poincar\'{e} sections (see Fig. \ref{fig:poincsecad}), we observed the difference in relative particle motion before and after the jump. 

\begin{figure}[h!]
\centering
\includegraphics[scale=0.43]{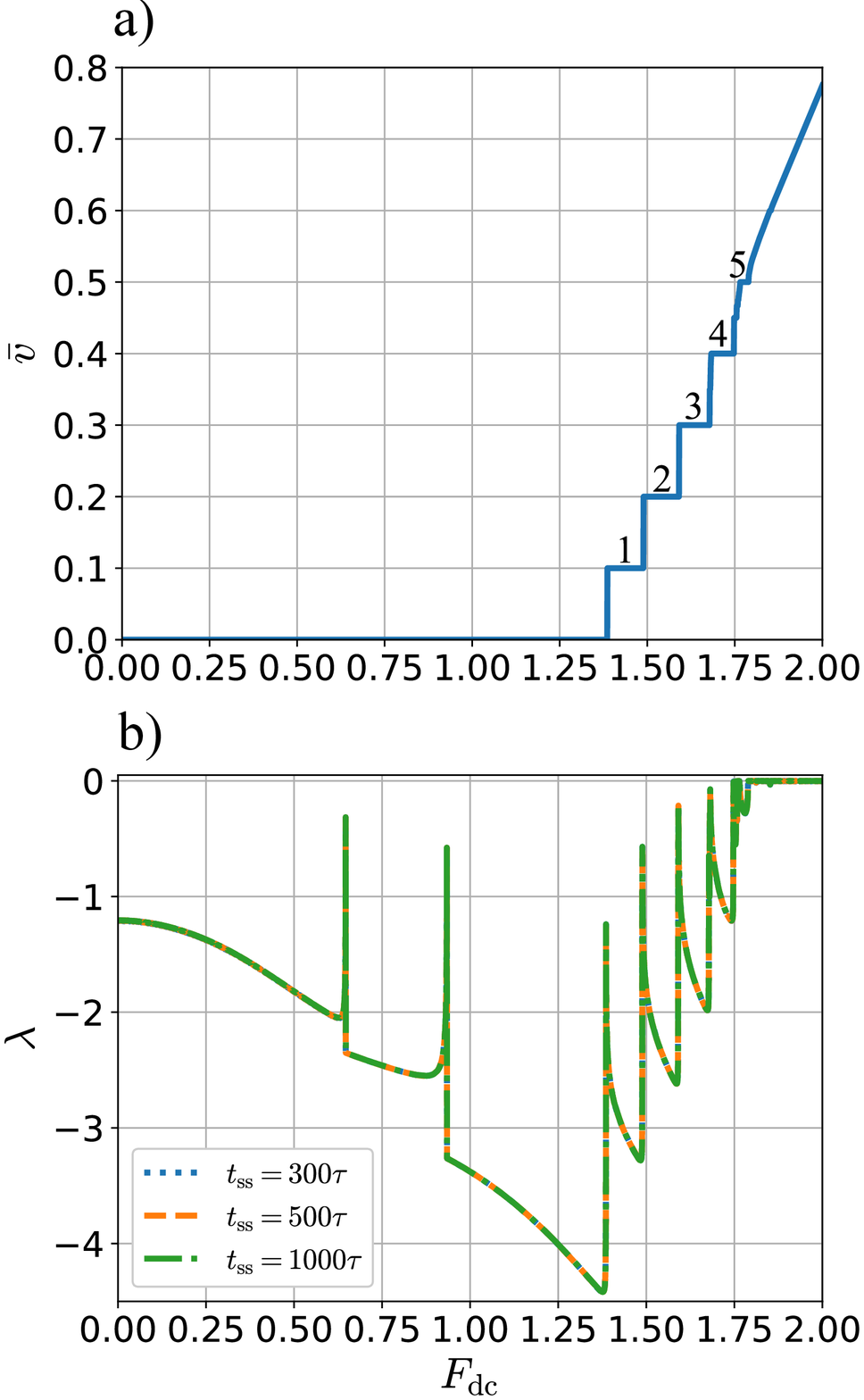} 
\caption{Average velocity a) and the largest LE b) as functions of driving force for $\omega=\frac{1}{2}$, $F_{\rm{ac}}=0.2$, $K=4.0$, $\nu_{0}=0.2$ and $r=0.7$. Large harmonic and subharmonic steps are marked.}
\label{r07}
\end{figure}

With further increasing of $r$ more jumps of the largest LE can be observed in the pinning regime (see, for instance, Fig. \ref{r07} for $r=0.7$, where two jumps emerge). This is a direct outcome of asymmetry of the successive narrow and wide neighboring substrate potential wells and thus in the case of standard FK model ($r=0$) these jumps are not observed \cite{application}. One can also see that the value of critical depinning force increased significantly for $r=0.7$. Such behavior is attributed to increasing discreteness and pinning of the system with increasing $r$ and critical force diverges as $r\to 1$ \cite{hu1}. However, there is a small region $0<|r|<0.1$ in which $F_{\rm{c}}$ decreases with the increase of shape parameter. In this area, the effect of changing oscillation frequencies of particles inside potential wells due to the increase of $r$, leading to more instability, becomes dominant and the result is lowering of the critical depinning force \cite{hu1}. This region of the decrease of critical force depends on the value of winding number as can be seen in \cite{saturation}. Also, the dynamics of the system in this area is quite similar to the standard FK model case and once again there are no jumps of the largest LE in the pinning regime.    

We also investigated the case of the integer values of winding number. Contrary to the standard case, subharmonic steps are detected for the generalized FK model with ASDP for integer values of $\omega$ \cite{hu1,farey}. However, the largest LE jumps in the pinning regime were not observed in this case. This should not be surprising as the interactions between the particles tend to make the system less stable and it is easier to displace it to another available configuration. For instance, in the case of $\omega=1$, there exists only one particle per a substrate potential well in average and this situation is much more robust to external disturbances than when there are two or more particles per a potential well in average. Consequently, the jumps of the largest LE in the pinning regime, originating from the structural changes in system's configuration, will not emerge.

\begin{figure}[h!]
\centering
\includegraphics[scale=0.45]{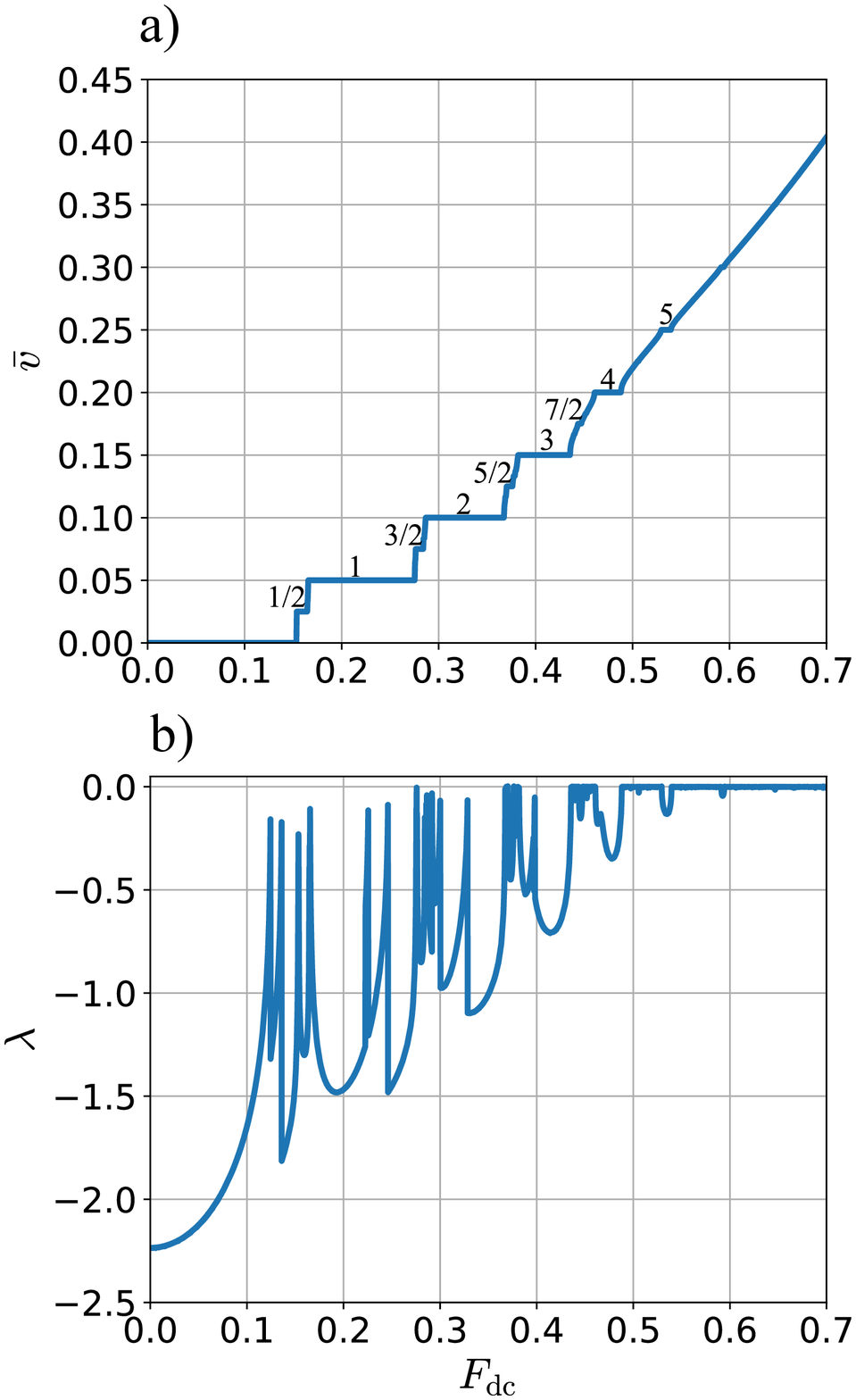}
\caption{Average velocity a) and the largest LE b) as functions of driving force for $\omega=\frac{1}{4}$, $F_{\rm{ac}}=0.2$, $K=4.0$, $\nu_{0}=0.2$ and $r=0.2$. Large harmonic and subharmonic steps are marked.}
\label{om14r02}
\end{figure}

On the other hand, more and more jumps of the largest LEs are expected to appear in the pinning regime as the average number of the particles per a substrate potential well increases, i.e. the rational value of winding number decreases. The response functions and the largest LEs were examined for multiple values of winding number $\omega=\frac{1}{q}$ and we have confirmed that the number of jumps increases with increasing $q$. As an example, in Fig. \ref{om14r02} one can observe two jumps in the pinning regime in case of $\omega=\frac{1}{4}$ and $r=0.2$, whereas for $\omega=\frac{1}{2}$ and $r=0.2$ only one jump in the pinning regime was present (see Fig \ref{fig:respfunler02}). 

Judging from the Figs. \ref{fig:respfunler02} and \ref{r07} one might intuitively presume that as the number of the jumps increases, the critical force rises as well. In this case, greater number of jumps and larger critical force are attributed to larger asymmetry of the successive substrate potential wells as the shape parameter rises. However, by comparing Figs. \ref{fig:respfunler02} and \ref{om14r02}, we see that greater number of jumps does not necessarily mean larger value of critical depinning force. In fact, for $\omega=\frac{1}{q}$ with increasing $q$, which also gives rise to the number of jumps of the largest LE in the pinning regime, the critical depinning force decreases before reaching its saturation value for larger shape parameters and diverging at $r\to 1$ \cite{saturation}. This is merely a consequence of the fact that the larger the average number of the particles per potential well is, the more unstable they are and easier to displace.  

\section{Conclusion}
\label{concl}

In this paper, dynamics of the driven overdamped generalized FK model with ASDP is investigated by analyzing corresponding response functions and the largest LEs. The obtained results have shown that the largest LE presents a more convenient method for analyzing the behavior of the system as it allows one to detect certain disturbances in the pinning regime. We confirmed that these disturbances of the largest LE in the pinning regime are attributed to the system's dynamics and not to the employed numerical methods or overdamped limit. Furthermore, the largest LE jumps in the pinning regime suggest that system goes through certain structural changes in chain configuration, which was confirmed by analyzing Poincar\'{e} sections for two neighboring particles. The number of these jumps increases with increasing the value of shape parameter and average number of particles per a substrate potential well since both of these factors contribute to lowering stability of the system. However, the number of the largest LE jumps in the pinning regime is not generally related to the critical depinning force or the behavior in the sliding regime due to existing saturation effects in these systems \cite{saturation}. Moreover, we showed that these jumps are also present in the sliding regime, where they are a consequence of a more complex structure of the largest LE on the step.

It is already widely known that the largest LE represents a reliable tool for characterizing chaotic, periodic and quasiperiodic motion. That is why it is important to outline that the present paper provides yet another useful purpose of the largest LE - spotting slight changes of the relative positions between the particles in the system. In other words, the largest LE analysis can be used not only for investigating integral quantities, that are accessible for experimental measurements, but also for detecting microchanges in chain configuration. Besides the discussed generalized FK model, the presented results can be applicable to any system of interacting oscillators under time dependent forces: the Fermi-Pasta-Ulam chain \cite{fpu}, Toda chain \cite{toda} and other related models. Furthermore, the observed jumps may appear in the charge density wave transport or in the irradiated Josephson junction systems with current driving and unharmonic current-phase relation in zero voltage state (static case). In the latter case, the LE jumps can be used as detectors of changes of phase differences, analogous to particle positions changes in the FK model, which are not observables in Josephson junction systems. Investigation of particle motion over a potential energy landscape is important for a deeper understanding of the physics of technologically important charge density wave systems, irradiated Josephson junctions, and vortex lattices. In such systems, typically, only averaged and integrated quantities are accessible for measurements and it is very difficult to investigate microscopic dynamics. Because of that, motion of colloidal particles in optical energy landscapes is studied with the purpose of observing the microscopic dynamics of particles in real space and time \cite{Nature,juniperpre,junipernjp}. The results from this manuscript can be experimentally tested by driving colloidal particles across an optical ASDP potential energy landscape.


\section{Appendix}
\label{app}

\begin{figure}[H]
\centering
\includegraphics[scale=0.73]{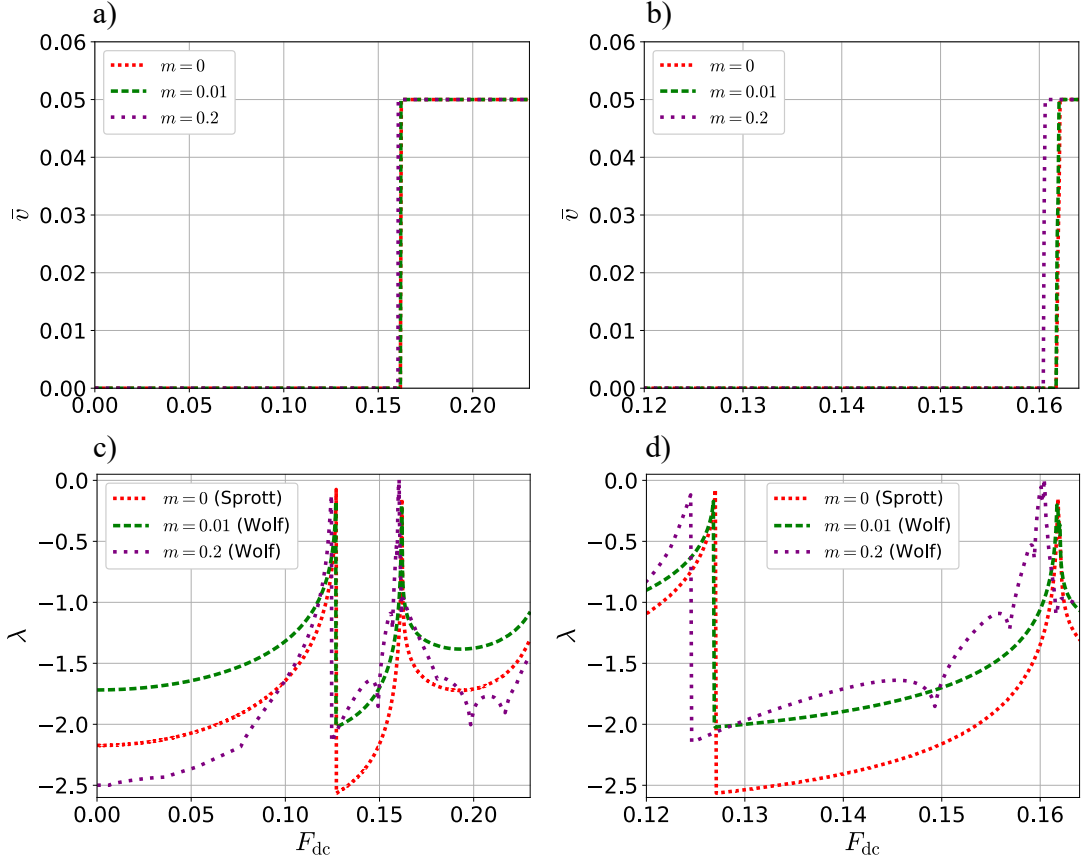}
\caption{Average velocity a) and the largest LE c) as functions of dc force for $\omega=\frac{1}{2}$, $F_{\rm{ac}}=0.2$, $K=4.0$, $\nu_{0}=0.2$, $r=0.2$ and three different values of mass $m$. Zoomed parts of a) and c) are shown in b) and d), respectively. The largest LE is calculated according to the Sprott's algorithm \cite{sprot} for the strictly overdamped limit $m=0$ whereas for two other masses we exploited Wolf's algorithm \cite{wolf}, which can be used to obtain full LE spectrum.} 
\label{masa}
\end{figure}

System of equations for the damped FK model is given by:
\begin{equation}
m \ddot{u}_{l}=u_{l+1}+u_{l-1}-2u_{l}-\dot{u}_{l}-V'(u_{l})+F(t), \hspace{4mm} l=1,2,...,N 
\end{equation}
where $m$ is particle mass. In Fig. \ref{masa} response function and the largest LE are shown for three different values of mass $m$. One can observe that, as $m\to 0$, dynamical behavior of the system becomes similar to the strictly overdamped model for $m=0$ (see equation \eqref{eqmot}). Furthermore, it is evident that the previously detected jumps of the largest LE in the pinning regime, which are attributed to the slight changes of the relative positions between the particles, are not a consequence of the strictly overdamped limit and all of the mentioned results are applicable to the damped FK model with inertial term as well.


\begin{acknowledgments}
The authors acknowledge financial support of the Ministry of Education, Science and Technological Development of the Republic of Serbia
 (Grant No. 451-03-47/2023-01/200125 and Grant No. 451-03-9/2021-14/200017).
\end{acknowledgments}



\end{document}